\documentclass{LT23auth}
\usepackage{graphicx}

\begin{document}

\begin{frontmatter}

% Use lower case letters in the title.
\title{The resonant magnetic mode: a common feature of  high-$T_C$
superconductors}

\author[address1]{Ph. Bourges\thanksref{thank1}},
\author[address2]{B. Keimer}
\author[address1]{S. Pailh\`es}
\author[address3]{L. P. Regnault}
\author[address1]{Y. Sidis}
\author[address2]{C. Ulrich}

\address[address1]{Laboratoire L\'{e}on Brillouin, CEA-CNRS, CE Saclay, 91191 Gif sur
Yvette, France}

\address[address2]{Max-Planck-Institut f\"{u}r Festk\"{o}rperforschung,
70569 Stuttgart, Germany}

\address[address3]{CEA Grenoble, D\'{e}partement de Recherche Fondamentale sur
la mati\`ere Condens\'ee, 38054~Grenoble cedex~9, France}

% The corresponding author should be distinguished and his email
% address and/or fax number must be given. His mailing address has to
% be complete: the proofs are send to this address around
% January 1, 2003. The address for sending proofs has to be indicated
% as "present address", if it is different from the address above.
\thanks[thank1]{Corresponding author.
 E-mail: bourges@llb.saclay.cea.fr}

\begin{abstract}
Inelastic neutron scattering experiments in high-$T_c$ cuprates
have evidenced a new magnetic excitation present in the
superconducting state. In particular, recent experiments on single
layer Tl$_{2}$Ba$_{2}$CuO$_{6+\delta }$,  performed  near optimum
doping ($ T_{c} \sim 90$ K), provide evidence of a sharp magnetic
resonant mode below $ T_{c}$, similar to previous reports on the
YBCO and BSCCO bilayer systems. This result supports models that
ascribe a key role to magnetic excitations in the mechanism of
superconductivity.
\end{abstract}

\end{frontmatter}

\section{Introduction}

Fifteen years after the discovery of high-$T_C$
superconductivity, its mechanism  is still the subject of vigorous debate. The
experimentally established $d_{x^2-y^2}$-wave symmetry of the
superconducting (SC) order parameter favors an origin based on antiferromagnetic (AF)
interactions \cite{scalapino}. However, a strong electron-phonon
coupling has been inferred from anomalies of the electronic
dispersion measured by photoemission spectroscopy experiments
\cite{shen-phonon}, among others. The resonant magnetic mode
observed in the inelastic neutron scattering (INS) spectra in the
SC state is one of the most persuasive experimental
indications of the importance of magnetic interactions for
superconductivity in the cuprates. Starting from the
YBa$_{2}$Cu$_{3}$O$_{6+x}$ (YBCO) system
\cite{rossat91,mook93,fong96,fong00,dai01}, numerous INS
 studies have established the existence of this
magnetic excitation mode as a generic feature of the high-T${\rm
_{c}}$ superconductors. The presence of the mode has been
demonstrated in another CuO$_{2}$-bilayer cuprate
Bi$_{2}$Sr$_{2}$CaCu$_{2}$O$_{8+\delta }$ (BSCCO)
\cite{fong99,he01} as well as in a single-layer system,
Tl$_{2}$Ba$_{2}$CuO$_{6+\delta }$ \cite{tl2201}. We here review
this feature for these different copper oxide systems where the
maximum superconducting temperature reaches 90 K.

\section{Resonance peak is generic to 90 K cuprates}

Since its first discovery in near-optimally doped YBCO
\cite{rossat91}, the resonance peak has been established as a
sharp excitation mode centered at the wave vector $(\pi,\pi)$ that
is characteristic of AF state in the insulating
cuprates. One of its key features is its disappearance in the
normal state: the peak intensity vanishes at T${\rm _{c}}$ without
substantial energy renormalization. Polarized neutron experiments
\cite{mook93,fong96}) have unambiguously demonstrated its magnetic
origin. For optimally doped YBCO and BSCCO, the peak is observed
close to $\sim$40 meV (table \ref{allreso}). One main
difference between the two systems is the broadening of the peak
in both energy and wave vector in BSCCO \cite{fong99}. In
particular, an energy width of $\sim$ 13 meV is found in BSCCO
whereas the peak is almost resolution-limited in YBCO. That
difference might be related to the intrinsic inhomogeneities found
in BSCCO by STM measurements where a substantial spatial distribution of the SC
gap was reported \cite{lang}. It is also known that the magnetic
resonance is extremely sensitive to defects: both magnetic (Ni)
\cite{sidisni} or non-magnetic (Zn) \cite{fong-zn} impurities
produce a marked energy broadening of the resonance peak in YBCO,
even in very dilute concentrations.

\begin{figure}[t]
%h=here, t=top, b=bottom, p=separate figure page
\begin{center}\leavevmode
\includegraphics[width=0.9\linewidth]{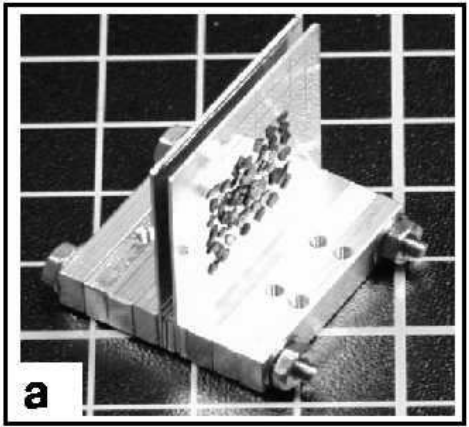}
\caption{ Photograph of the array of co-oriented
Tl$_{2}$Ba$_{2}$CuO$_{6+\delta }$ single crystals. The crystals
are glued onto Al plates only two of which are shown for clarity.
The entire multicrystal array is equivalent to a single crystal
with a mosaicity of 1.5$^\circ$ on which regular transverse
acoustic phonons can be measured (from
\cite{tl2201}).}\label{tlmount}\end{center}\end{figure}

Most electronic properties of the high-T${\rm _{c}}$
superconductors (including superconductivity) are assumed to be
determined by the commonly shared $\rm CuO_{2}$ layers. As a
result, most theoretical models of high temperature
superconductivity are based on a two-dimensional square lattice.
As the resonant mode had only been observed in copper oxide
systems which  have in common two $\rm CuO_{2}$ layers per unit
cell, the relevance of this mode for all high-T${\rm _{c}}$
superconductors was questioned, as it could be related to
interlayer interactions that vary substantially among the copper
oxides. In particular, the resonance peak has never been reported
in the most widely studied single layer cuprate, ${\rm
La_{2-x}Sr_xCuO_{4}}$. However, the maximum T${\rm
_{c}}$ of this system is only about 38 K. Until recently, the
investigation of another single layer cuprate whose maximum T${\rm
_{c}}$ can reach 90 K remained one of the major challenges in the
field.

\begin{figure}[t]
%h=here, t=top, b=bottom, p=separate figure page
\begin{center}\leavevmode
\includegraphics[width=0.9\linewidth,height=6 cm]{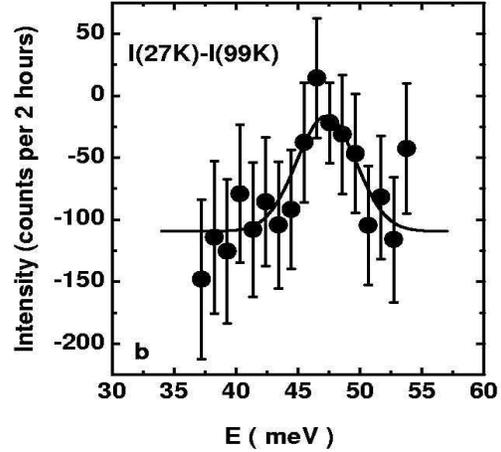}
\caption{ Difference between constant-{\bf Q} scans measured
at $T=99$ K ($\rm > T_c$) and $T=27$ K ($\rm < T_c$) at the
antiferromagnetic wavevector (from \cite{tl2201}).}\label{tlw}\end{center}
\end{figure}

We therefore recently focused our effort on
Tl$_{2}$Ba$_{2}$CuO$_{6+\delta }$, a material with unbuckled,
widely spaced $ \rm CuO_{2}$ layers and a maximum T${\rm _{c}}$
around 90 K. The crystal growth of the Tl-based copper oxide
superconductors suffers from technical difficulties arising from
the toxicity of Tl, and only single crystals with moderate volumes
of about $\sim$ 0.5-3 mm$^{3}$ can be obtained through a CuO-rich
flux technique \cite{kolesnikov}. To perform INS, which requires
large single crystals of minimum volume $\sim$0.1 cm$^{3}$, an
array of more than 300 co-aligned single crystals had to be
assembled (see photo Fig.\ref{tlmount}) \cite{tl2201}. Using
high-flux triple axis spectrometers at LLB-Saclay (France) and
ILL-Grenoble (France), we were able to observe a resonance peak
for this array \cite{tl2201}. Fig. \ref{tlw} shows the difference
of energy scans in the SC state and the normal state
at the AF wave vector: it exhibits the
characteristic signature of the resonant mode, albeit at an energy
of 47 meV that is somewhat larger than the mode energy in the
bilayer compounds (see table \ref{allreso}).

In many aspects, the resonant mode in
Tl$_{2}$Ba$_{2}$CuO$_{6+\delta }$ shows strong similarities with
that observed in YBCO$_7$. In both cases, the mode is limited by
the resolution in energy and exhibits the same extension in
$q$-space (see table \ref{allreso}). Further, the resonance peak
spectral weight per ${\rm CuO_2}$ layer, defined as $\int d\omega
d^3{\bf Q} Im \chi^{res}({\bf Q},\omega)$, equals $(0.02
\mu_B^2/eV)$ for both of these systems. In BSCCO, where the
resonance $q$-width along the diagonal (110) direction is about
twice as large, the spectral weight is larger, but the energy
integrated intensity at the AF wavevector,
normalized to one ${\rm CuO_2}$ layer, is almost identical for the
three different cuprates (see table \ref{allreso}).

\section{Resonance peak spectral weight}

\begin{table}[t]
\begin{center}
\begin{tabular}{| c | c c | c|}
\hline
& 2-layers & & 1-layer \\
 \hline
Cuprates & YBCO  & BSCCO & Tl-2201 \\
 \hline
Resonance energy (meV) & 41 & 43  & 47 \\
$E_r/k_B T_c$ & 5.1 & 5.4 & 6 \\
\hline
 $\Delta_q$(\AA$^{-1}$) & 0.25 &  0.52 & 0.23 \\
\hline
$\int d\omega Im \chi^{res}((\pi,\pi),\omega) (\mu_B^2/eV f.u.)$ & 1.6 & 1.9 & 0.7 \\
$\int d\omega d^3{\bf Q} Im \chi^{res}{\bf Q},\omega) (\mu_B^2/eV f.u.)$ & 0.043 & 0.23 & 0.02 \\
\hline
\end{tabular}
\caption{Resonance peak energies, $q$-width (Full Width at Half
Maximum) and spectral weights per formula unit at optimal doping
for three different cuprates. }\label{allreso}
\end{center}
\end{table}

\begin{figure}[t]
%h=here, t=top, b=bottom, p=separate figure page
\begin{center}\leavevmode
\includegraphics[width=0.98\linewidth]{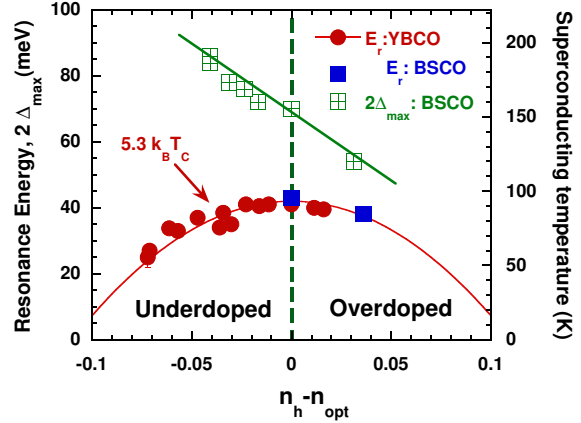}
\caption{ Doping dependence of the resonance energy at $(\pi,\pi)$
in YBCO and BSCCO  from Refs.
\cite{rossat91,fong00,dai01,fong99,he01} as a function of various
doping levels referenced to the optimal doping level, $n_{opt}$,
corresponding to $T_C^{max}$. The doping level has been determined
through the empirical relation
$T_C/T_C^{max}=1-82.6(n_h-n_{opt})^2$\cite{tallon}. $n_{opt}$ is
generally identified as a hole doping level of 16 \% although this
assumption is not necessary here. The red full curve shows the
doping dependence of the superconducting temperature times $5.3$.
The figure also shows twice the maximum SC gap as measured in
BSCCO by ARPES \cite{mesot}.
}\label{aller}\end{center}\end{figure}

The resonant mode at $(\pi,\pi)$ represents the major part of the
experimentally observable magnetic spectrum for the optimally
doped cuprates. Going into the underdoped state, as experimentally
realized for different oxygen contents in YBCO
\cite{fong00,dai01}, this mode shifts to lower energy as shown in
Fig. \ref{allreso}. Further, INS studies of underdoped
YBa$_2$Cu$_3$O$_{6+x}$ reveal a complex lineshape incorporating
the resonant mode at $(\pi,\pi)$. Across T${\rm _{c}}$, a drastic
magnetic intensity redistribution occurs in both momentum space
\cite{science} and energy \cite{fong00}. Incommensurate
excitations at somewhat lower energies than the resonance peak
energy at $(\pi,\pi)$ are observed. They form a downward
dispersion relation continuously connected to the commensurate
resonance peak energy at $(\pi,\pi)$ \cite{science}. In the
overdoped regime, as observed so far only in one  BSCCO sample
\cite{he01}, the resonant mode also shifts down to lower energy
(Fig. \ref{aller}).

Besides the observation of the magnetic resonance peak below T${\rm _{c}}$ by INS,
anomalies in the quasiparticle spectra has been also reported by photoemission \cite{infavor}, optical
conductivity \cite{carbotte}, tunneling \cite{zaza}, and Raman
scattering techniques. Their interpretation as an evidence of the
coupling of quasiparticules to the neutron mode has
stimulated spin fluctuation based pairing scenarios which are,
however, still controversial \cite{shen-phonon}.
It should be
noted that, for all doping levels in YBCO, the resonance peak
spectral weight is almost constant \cite{fong00}, $ \int d\omega
d^3{\bf Q} Im \chi^{res}{\bf Q},\omega)  \simeq$ 0.05
$\mu_B^2/f.u.$, and represents about 2 \% of the spectral weight
contained in the spin wave spectrum of undoped YBCO$_6$. This
seemingly rather small spectral weight is, however, highly
concentrated at wave vectors that connect extended saddle points
in the band structure.  Depending on how the magnetic spectrum and
the density of states of charged quasiparticles are modeled, the
resonance peak cannot \cite{kka} or can \cite{abanov02} account
for various anomalies detected in charge spectroscopies of the
cuprates. In any case, the spectral weight is large enough to
account for a sizable fraction of the superconducting condensation
energy \cite{kka,abanov02,ec2}, and so the resonant mode must be
considered as a key player in theories of high-T${\rm _{c}}$
superconductivity.

\section{Models}

In any superconductors, INS experiments have the potential to perform a complete identification of 
the symmetry of superconducting order parameter. As emphasized by Joynt and Rice\cite{joynt}, 
the knowledge of wavevector and energy-dependent  spin susceptibility in superconductors
indeed reflects directly the vector structure of the SC gap function.
In high-T${\rm _{c}}$ superconductors,
the observation magnetic excitations at $(\pi,\pi)$ 
is, for instance, fully consistent with the d-wave symmetry of SC order parameter\cite{fong95}.

% In conventional superconductors, it has been proposed for many decade that
% residual repulsive interaction could lead to the formation of low
% energy particle-hole bound states. These excitons have never been observed so far.
More precisely, the resonant mode can be theoretically modeled as a spin exciton
collective bound state pulled  below the threshold
 of particle-hole (p-h) excitation continuum in the $d$-wave SC state by magnetic interactions
\cite{models,bl}. The consistency of this idea can be quantitatively tested for the BSCCO system where the
SC gap and the Fermi surface have been measured by
photoemission spectroscopy. A close inspection of the Fermi
surface \cite{fermisurface} shows that the quasi-particles which
are connected by the $(\pi,\pi)$ momentum (corresponding to the
"hot spots") have the energy $E_k \simeq 0.9 \Delta_{max}$ at the
Fermi level. The p-h continuum threshold would then be
$\omega_c \simeq 1.8 \Delta_{max}  \simeq 63$ meV ( where the SC
gap measured at optimal doping is $\Delta_{max}$= 35 meV
\cite{mesot}). The resonance peak is observed at optimal doping at
an energy around 43 meV, clearly lower than $\omega_c$. A ratio
$E_r \simeq 1.2 \Delta_{max}$ is found, indicating that the
resonance peak occurs well below the p-h continuum in
agreement with this theoretical approach. To confirm these ideas,
one might eventually be able to detect the p-h spin-flip
continuum threshold directly by INS. For YBCO, BSCCO and
Tl$_{2}$Ba$_{2}$CuO$_{6+\delta }$, $2 \Delta_{max}$ can also be
determined by the position of the B$_{\rm 1g}$ mode in Raman
scattering, which occurs around 70 meV for the three cuprates
\cite{raman}. Furthermore, the observation of a higher resonance
energy in Tl$_{2}$Ba$_{2}$CuO$_{6+\delta }$ (whereas the SC gap
maximum is similar) can be explained by the non-negligible value
of the interplane magnetic coupling in bilayer systems which could
push the bound state to lower energy.
Therefore, the spin exciton
scenario is consistent for the different cuprates assuming a
similar Fermi surface topology. As shown by Fig. \ref{aller}, the
resonance peak energy exhibits a dome-like shape as a function of
doping following, an approximately linear relationship with T${\rm
_{c}}$ \cite{he01} on both sides of the optimal doping level.

Alternatively, other approaches \cite{morr,sachdev} associate the resonance peak to a pre-existing
soft mode reminiscent of nearby (commensurate or incommensurate)
AF phase. While the collective mode decays into p-h excitations in the normal state, 
it becomes undamped below $\omega_c$ in the SC state, giving rise to neutron peak.
In that framework, the apparent absence of the mode in the ${\rm La_{2-x}Sr_xCuO_{4}}$ system,
could imply that the condition $E_r < \omega_c$ may not be satisfied
\cite{morr}.In addition, predictions of the influence of
impurities on the line shape of the mode \cite{sachdev} agree well
with measurements in YBCO \cite{fong-zn}. 

In line with those approaches, the dispersive resonance peak  has been re-examined
within a stripe model \cite{batista}, where
the observed dispersion \cite{science} corresponds to the trace of
spin-waves emanating from incommensurate Bragg reflections shifted
away from ($\pi,\pi$). This approach agrees with the spin dynamics
measured in stripe-ordered nickelates \cite{lsno}. However, a
detailed comparison of the temperature and momentum dependences of
the magnetic spectrum in the cuprates and nickelates does not
favor this model as a general scenario for the cuprates.

\section{Conclusion}

In conclusion, the magnetic resonance is a generic feature of
high-T${\rm _{c}}$ cuprates, at least for systems with a maximum
$T_C^{max}$ of around 90 K. The resonant mode occurs at an energy
$E_r$ always lower than twice the superconducting gap (Fig.
\ref{aller}) ${ E_r < 2 \Delta_{max} \sim}$ 70 meV $\sim 9 k_B
T_c$, which has been deduced either by photoemission measurements
in BSCCO \cite{mesot} or by the position of the B$_{\rm 1g}$ mode
in Raman scattering; the latter has been measured for all
cuprates. This agrees with models %\cite{models,morr,sachdev} 
which interpret the resonant mode as a magnetic collective mode of the
d$_{x^2-y^2}$-wave superconducting state below the electron-hole
continuum.

\end{document}